\documentclass[11pt]{article}
\usepackage{amssymb} 
\usepackage{amsfonts} 
\usepackage{amsthm}
\usepackage{latexsym}
\usepackage[T1]{fontenc}
\usepackage{times}
\DeclareMathAlphabet{\mathib}{T1}{ptm}{b}{it}

\newtheorem{thm}{Theorem}
\newtheorem*{thm*}{Theorem}
\newtheorem*{lemma*}{Lemma}
\newtheorem{lemma}{Lemma}
\theoremstyle{remark}
\newtheorem{example}{Example}
\newcommand{\abs}[1]{\mbox{$|#1|$}}
\newcommand{\locsys}{(\mathcal{H},\Delta \mapsto
  E_{\Delta},\mathbf{a}\mapsto U(\mathbf{a}))}
\newcommand{\norm}[1]{\mbox{$\| #1\|$}}
\newcommand{\hil}[1]{\mathcal{#1}}
\title{No place for particles in relativistic quantum theories?}
\author{Hans Halvorson and Rob Clifton \\
{\small {\it Department of Philosophy, University of Pittsburgh }} \\
{\small {\it hphst1@pitt.edu, rclifton@pitt.edu }}}
\date{}
\begin{document}
\maketitle 
\begin{abstract} Several recent arguments purport to show that 
  there can be no relativistic, quantum-mechanical theory of
  localizable particles and, thus, that relativity and quantum
  mechanics can be reconciled only in the context of quantum field
  theory.  We point out some loopholes in the existing arguments, and
  we provide two no-go theorems to close these loopholes.  However,
  even with these loopholes closed, it does not yet follow that
  relativity plus quantum mechanics exclusively requires a field
  ontology, since relativistic quantum field theory itself might
  permit an ontology of localizable particles supervenient on the
  fundamental fields.  Thus, we provide another no-go theorem to rule
  out this possibility.  Finally, we allay potential worries about
  this conclusion by arguing that relativistic quantum field theory
  can nevertheless explain the possibility of ``particle detections'',
  as well as the pragmatic utility of ``particle talk.''
\end{abstract}

\section{Introduction}
It is a widespread belief, at least within the physics community, that
there is no particle mechanics that is simultaneously relativistic and
quantum-theoretic; and, thus, that the only relativistic quantum
theory is a \emph{field} theory.  This belief has received much
support in recent years in the form of rigorous ``no-go theorems'' by
Malament (1996) and Hegerfeldt (1998a, 1998b).  In particular,
Hegerfeldt shows that in a generic quantum theory (relativistic or
non-relativistic), if there are states with localized particles, and
if there is a lower bound on the system's energy, then superluminal
spreading of the wavefunction must occur.  Similarly, Malament shows
the inconsistency of a few intuitive desiderata for a relativistic,
quantum-mechanical theory of (localizable) particles.  Thus, it
appears that there is a fundamental conflict between the demands of
relativistic causality and the requirements of a theory of localizable
particles.

What is the philosophical lesson of this apparent conflict between
relativistic causality and localizability?  One the one hand, if we
believe that the assumptions of Malament's theorem must hold for any
theory that is descriptive of our world, then it follows that our
world cannot be correctly described by a particle theory.  On the
other hand, if we believe that our world \emph{can} be correctly
described by a particle theory, then one (or more) of the Malament's
assumptions must be false.  Malament clearly endorses the first
response; that is, he argues that his theorem entails that there is no
relativistic quantum mechanics of localizable particles (insofar as
any relativistic theory precludes act-outcome correlations at
spacelike separation).  Others, however, have argued that the
assumptions of Malament's theorem need not hold for any relativistic,
quantum-mechanical theory (cf.~Fleming and Butterfield 1999), or that
we cannot judge the truth of the assumptions until we resolve the
interpretive issues of elementary quantum mechanics (cf.~Barrett
2000).

Although we do not think that these arguments against Malament's
assumptions succeed, there are other reasons to doubt that Malament's
theorem is sufficient to support a sound argument against the
possibility of a relativistic quantum mechanics of localizable
particles.  First, Malament's theorem depends on a specific assumption
about the structure of Minkowski spacetime---a ``no preferred
reference frame'' assumption---that could be seen as having less than
full empirical warrant.  Second, Malament's theorem establishes only
that there is no relativistic quantum mechanics in which particles can
be completely localized in spatial regions with sharp boundaries; it
leaves open the possibility that there might be a relativistic quantum
mechanics of ``unsharply'' localized particles.  In this paper, we
present two new no-go theorems which, together, suffice to close these
loopholes in the argument against relativistic quantum mechanics.
First, we present a strengthened no-go theorem that subsumes the
results of Malament and Hegerfeldt, and which does not depend on the
``no preferred frame'' assumption (Theorem~\ref{improve}).  Second, we
derive a generalized version of Malament's theorem that shows that
there is no relativistic quantum mechanics of ``unsharply'' localized
particles (Theorem~\ref{pov}).

However, it would be a mistake to think that these result show---or,
are intended to show---that a field ontology, rather than a particle
ontology, is appropriate for relativistic quantum theories.  While
these results show that there are no position observables that satisfy
certain relativistic constraints, quantum field theories---both
relativistic \emph{and} non-relativistic---already reject the notion
of position observables in favor of ``localized'' field observables.
Thus, no-go results against relativistic position operators have
nothing to say about the possibility that relativistic quantum field
theory might permit a ``particle interpretation,'' in which localized
particles are supervenient on the underlying localized field
observables.  To exclude this latter possibility, we formulate (in
Section~6) a necessary condition for a generic quantum theory to
permit a particle interpretation, and we then show that this condition
fails in \emph{any} relativistic theory (Theorem~\ref{rqft}).

Since \emph{our world} is presumably both relativistic and
quantum-theoretic, these results show that there are no localizable
particles.  However, in Section~7 we shall argue that relativistic
quantum field theory itself warrants an approximate use of ``particle
talk'' that is sufficient to save the phenomena.

\section{Malament's Theorem}
Malament's theorem shows the inconsistency of a few intuitive
desiderata for a relativistic quantum mechanics of (localizable)
particles.  It strengthens previous results (e.g., Schlieder 1971) by
showing that the assumption of ``no superluminal wavepacket
spreading'' can be replaced by the weaker assumption of
``microcausality,'' and by making it clear that Lorentz invariance is
not needed to derive a conflict between relativistic causality and
localizability.

In order to present Malament's result, we assume that our background
spacetime $M$ is an affine space, with a foliation $\mathcal{S}$ into
spatial hyperplanes.  (For ease, we can think of an affine space as a
vector space, so long as we do not assign any physical significance to
the origin.)  This will permit us to consider a wide range of
relativistic (e.g., Minkowski) as well as non-relativistic (e.g.,
Galilean) spacetimes.  The pure states of our quantum-mechanical
system are given by rays in some Hilbert space $\hil{H}$.  We assume
that there is a mapping $\Delta \mapsto E_{\Delta}$ of \emph{bounded}
subsets of hyperplanes in $M$ into projections on $\hil{H}$.  We think
of $E_{\Delta}$ as representing the proposition that the particle is
localized in $\Delta$; or, from a more operational point of view,
$E_{\Delta}$ represents the proposition that a position measurement is
certain to find the particle within $\Delta$.  We also assume that
there is a strongly continuous representation $\mathbf{a}\mapsto
U(\mathbf{a})$ of the translation group of $M$ in the unitary
operators on $\hil{H}$.  Here strong continuity means that for any
unit vector $\psi \in \hil{H}$, $\langle \psi,U(\mathbf{a})\psi
\rangle \rightarrow 1$ as $\mathbf{a}\rightarrow 0$; and it is
equivalent (via Stone's theorem) to the assumption that there are
energy and momentum observables for the particle.  If all of the
preceding conditions hold, we say that the triple $\locsys$ is a
\emph{localization system} over $M$.

The following conditions should hold for any localization
system---either relativistic or non-relativistic---that describes a
single particle.
\begin{description}
\item[{\it Localizability:}] If $\Delta $ and $\Delta '$ are disjoint
  subsets of a single hyperplane, then \newline
  \mbox{$E_{\Delta}E_{\Delta '}=0$}.
\item[{\it Translation covariance:}] For any $\Delta$ and for any
  translation $\mathbf{a}$ of $M$, \newline
  $U(\mathbf{a})E_{\Delta}U(\mathbf{a})^{*}=E_{\Delta +\mathbf{a}}$.
\item[{\it Energy bounded below:}] For any timelike translation
  $\mathbf{a}$ of $M$, the generator $H(\mathbf{a})$ of the
  one-parameter group $\{ U(t\mathbf{a}):t\in \mathbb{R} \}$ has a
  spectrum bounded from below. \end{description} We recall briefly the
motivation for each of these conditions.  ``Localizability'' says that
the particle cannot be detected in two disjoint spatial sets at a
given time.  ``Translation covariance'' gives us a connection between
the symmetries of the spacetime $M$ and the symmetries of the
quantum-mechanical system.  In particular, if we displace the particle
by a spatial translation $\mathbf{a}$, then the original wavefunction
$\psi$ will transform to some wavefunction $\psi _{\mathbf{a}}$.
Since the statistics for the displaced detection experiment should be
identical to the original statistics, we have $\langle \psi
,E_{\Delta}\psi \rangle =\langle \psi _{\mathbf{a}},E_{\Delta
  +\mathbf{a}}\psi _{\mathbf{a}}\rangle$.  By Wigner's theorem,
however, the symmetry is implemented by some unitary operator
$U(\mathbf{a})$.  Thus, $U(\mathbf{a})\psi =\psi _{\mathbf{a}}$, and
$U(\mathbf{a})E_{\Delta}U(\mathbf{a})^{*}=E_{\Delta +\mathbf{a}}$.  In
the case of time translations, the covariance condition entails that
the particle has unitary dynamics.  (This might seem to beg the
question against a collapse interpretation of quantum mechanics; we
dispell this worry at the end of this section.)  Finally, the ``energy
bounded below'' condition asserts that, relative to any free-falling
observer, the particle has a lowest possible energy state.  If it were
to fail, we could extract an arbitrarily large amount of energy from
the particle as it drops down through lower and lower states of
energy.

We now turn to the ``specifically relativistic'' assumptions needed
for Malament's theorem.  The special theory of relativity entails that
there is a finite upper bound on the speed at which (detectable)
physical disturbances can propagate through space.  Thus, if $\Delta$
and $\Delta '$ are distant regions of space, then there is a positive
lower bound on the amount of time it should take for a particle
localized in $\Delta$ to travel to $\Delta'$.  We can formulate this
requirement precisely by saying that for any timelike translation
$\mathbf{a}$, there is an $\epsilon >0$ such that, for every state
$\psi$, if $\langle \psi ,E_{\Delta}\psi \rangle =1$ then $\langle
\psi ,E_{\Delta '+t\mathbf{a}}\psi \rangle =0$ whenever $0\leq
t<\epsilon$.  This is equivalent to the following assumption.
\begin{description}
\item[{\it Strong causality:}] If $\Delta$ and $\Delta '$ are disjoint
  subsets of a single hyperplane, and if the distance between $\Delta$
  and $\Delta '$ is nonzero, then for any timelike translation
  $\mathbf{a}$, there is an $\epsilon >0$ such that
  $E_{\Delta}E_{\Delta '+t\mathbf{a}}=0$ whenever $0\leq t<\epsilon$.
\end{description} (Note that strong causality entails localizability.)
Although strong causality is a reasonable condition for relativistic
theories, Malament's theorem requires only the following weaker
assumption (which he himself calls ``locality'').
\begin{description}
\item[{\it Microcausality:}] If $\Delta$ and $\Delta '$ are disjoint
  subsets of a single hyperplane, and if the distance between $\Delta$
  and $\Delta '$ is nonzero, then for any timelike translation
  $\mathbf{a}$, there is an $\epsilon >0$ such that
  $[E_{\Delta},E_{\Delta '+t\mathbf{a}}]=0$ whenever $0\leq
  t<\epsilon$.  \end{description} If $E_{\Delta}$ can be measured
within $\Delta$, microcausality is equivalent to the assumption that a
measurement within $\Delta$ cannot influence the statistics of
measurements performed in regions that are spacelike to $\Delta$ (see
Malament 1996, 5).  Conversely, a failure of microcausality would
entail the possibility of act-outcome correlations at spacelike
separation.  Note that both strong and weak causality make sense for
non-relativistic spacetimes (as well as for relativistic spacetimes);
though, of course, we should not expect either causality condition to
hold in the non-relativistic case.

\begin{thm*}[Malament] Let $(\hil{H},\Delta \mapsto
  E_{\Delta},\mathbf{a}\mapsto U(\mathbf{a}))$ be a localization
  system over Minkowski spacetime that satisfies:
\begin{enumerate} 
\item Localizability
\item Translation covariance
\item Energy bounded below
\item Microcausality
\end{enumerate} Then $E_{\Delta}=0$ for all
$\Delta$.  \label{malament} \end{thm*}
\noindent Thus, in every state, there is no chance that the particle
will be detected in any local region of space.  As Malament claims,
this serves as a \emph{reductio ad absurdum} of any relativistic
quantum mechanics of a single (localizable) particle.

Several authors have claimed that Malament's theorem is not sufficient
to rule out a relativistic quantum mechanics of localizable particles.
In particular, these authors argue that it is not reasonable to expect
the conditions of Malament's theorem to hold for any relativistic,
quantum-mechanical theory of particles.  For example, Dickson (1997)
argues that a `quantum' theory does not need a position
\emph{operator} (equivalently, a system of localizing projections) in
order to treat position as a physical quantity; Barrett (2000) argues
that time-translation covariance is suspect; and Fleming and
Butterfield (1999) argue that the microcausality assumption is not
warranted by special relativity.  We now show, however, that none of
these arguments is decisive against the assumptions of Malament's
theorem.

Dickson (1997, 214) cites the Bohmian interpretation of the Dirac
equation as a counterexample to the claim that any `quantum' theory
must represent position by an operator.  In order to see what Dickson
might mean by this, recall that the Dirac equation admits both
positive and negative energy solutions.  If $\hil{H}$ denotes the
Hilbert space of all (both positive and negative energy) solutions,
then we may define the `standard position operator' $Q$ by setting
$Q\psi (\mathib{x})=\mathib{x}\cdot \psi (\mathib{x})$ (Thaller 1992,
7).  If, however, we restrict to the Hilbert space
$\hil{H}_{\mathrm{pos}}\subset \hil{H}$ of positive energy solutions,
then the probability density given by the Dirac wavefunction does not
correspond to a self-adjoint position operator (Thaller 1992, 32).
According to Holland (1993, 502), this lack of a position operator on
$\hil{H}_{\mathrm{pos}}$ precludes a Bohmian interpretation of $\psi
(\mathib{x})$ as a probability amplitude for finding the particle in
an elementary volume $d^{3}\mathib{x}$ around $\mathib{x}$.

Since the Bohmian interpretation of the Dirac equation uses all states
(both positive and negative energy), and the corresponding position
observable $Q$, it is not clear what Dickson means by saying that the
Bohmian interpretation of the Dirac equation dispenses with a position
observable.  Moreover, since the energy is not bounded below in
$\hil{H}$, this would not in any case give us a counterexample to
Malament's theorem.  However, Dickson could have developed his
argument by appealing to the positive energy subspace
$\hil{H}_{\mathrm{pos}}$.  In this case, we \emph{can} talk about
positions despite the fact that we do not have a position observable
in the usual sense.  In particular, we shall show in
Section~\ref{unsharp} that, for talk about positions, it suffices to
have a family of ``unsharp'' localization observables.  (And, yet, we
shall show that relativistic quantum theories do not permit even this
attenuated notion of localization.)

Barrett (2000) argues that the significance of Malament's theorem
cannot be assessed until we have solved the measurement problem:
\begin{quote} If we might have to violate
  the apparently weak and obvious assumptions that go into proving
  Malament's theorem in order to get a satisfactory solution to the
  measurement problem, then all bets are off concerning the
  applicability of the theorem to the detectible entities that inhabit
  our world.  (Barrett 2000, 16) \end{quote} In particular, a solution
to the measurement problem may require that we abandon unitary
dynamics.  But if we abandon unitary dynamics, then the translation
covariance condition does not hold, and we need not accept the
conclusion that there is no relativistic quantum mechanics of
(localizable) particles.

Unfortunately, it is not clear that we could avoid the upshot of
Malament's theorem by moving to a collapse theory.  Existing
(non-relativistic) collapse theories take the empirical predictions of
quantum theory seriously.  That is, the ``statistical algorithm'' of
quantum mechanics is assumed to be at least approximately correct; and
collapse is introduced only to ensure that we obtain determinate
properties at the end of a measurement.  However, in the present case,
Malament's theorem shows that the statistical algorithm of any quantum
theory predicts that if there are local particle detections, then
act-outcome correlations are possible at spacelike separation.  Thus,
if a collapse theory is to stay close to these predictions, it too
would face a conflict between localizability and relativistic
causality.

Perhaps, then, Barrett is suggesting that the price of accomodating
localizable particles might be a complete abandonment of unitary
dynamics, \emph{even at the level of a single particle}.  In other
words, we may be forced to adopt a collapse theory \emph{without}
having any underlying (unitary) quantum theory.  But even if this is
correct, it wouldn't count against Malament's theorem, which was
intended to show that there is no relativistic \emph{quantum} theory
of localizable particles.  Furthermore, noting that Malament's theorem
requires unitary dynamics is one thing; it would be quite another
thing to provide a model in which there \emph{are} localizable
particles---at the price of non-unitary dynamics---but which is also
capable of reproducing the well-confirmed quantum interference effects
at the micro-level.  Until we have such a model, pinning our hopes for
localizable particles on a failure of unitary dynamics is little more
than wishful thinking.

Like Barrett, Fleming (Fleming and Butterfield 1999, 158ff) disagrees
with the reasonableness of Malament's assumptions.  Unlike Barrett,
however, Fleming provides a concrete model in which there are
localizable particles (viz., using the Newton-Wigner position operator
as a localizing observable) and in which Malament's microcausality
assumption fails.  Nonetheless, Fleming argues that this failure of
microcausality is perfectly consistent with relativistic causality.

According to Fleming, the property ``localized in $\Delta$''
(represented by $E_{\Delta}$) need not be detectable within $\Delta$.
As a result, $[E_{\Delta},E_{\Delta '}]\neq 0$ does not entail that it
is possible to send a signal from $\Delta$ to $\Delta '$.  However, by
claiming that local \emph{beables} need not be local
\emph{observables}, Fleming undercuts the primary utility of the
notion of localization, which is to indicate those physical quantities
that are operationally accessible in a given region of spacetime.
Indeed, it is not clear what motivation there could be---aside from
indicating what is locally measurable---for assigning observables to
spatial regions.  If $E_{\Delta}$ is \emph{not} measurable in
$\Delta$, then why should we say that ``$E_{\Delta}$ is localized in
$\Delta$''?  Why not say instead that ``$E_{\Delta}$ is localized in
$\Delta '$'' (where $\Delta '\neq \Delta$)?  Does either statement
have any empirical consequences and, if so, how do their empirical
consequences differ?  Until these questions are answered, we maintain
that local beables are always local observables; and a failure of
microcausality \emph{would} entail the possibility of act-outcome
correlations at spacelike separation.  Therefore, the microcausality
assumption is an essential feature of any relativistic quantum theory
with ``localized'' observables.  (For a more detailed argument along
these lines, see Halvorson 2001, Section 6.)

Thus, the arguments against the four (explicit) assumptions of
Malament's theorem are unsuccessful; these assumptions are perfectly
reasonable, and we should expect them to hold for any relativistic,
quantum-mechanical theory.  However, there is another difficulty with
the argument against any relativistic quantum mechanics of
(localizable) particles: Malament's theorem makes \emph{tacit} use of
specific features of Minkowski spacetime which---some might
claim---have less than perfect empirical support.  First, the
following example shows that Malament's theorem fails if there is a
preferred reference frame.

\begin{example} \label{aristotle} Let $M=\mathbb{R}^{1}\oplus
  \mathbb{R}^{3}$ be full Newtonian spacetime (with a distinguished
  timelike direction $\mathbf{a}$).  To any set of the form $\{
  (t,x):x\in \Delta \}$, with $t\in \mathbb{R}$, and $\Delta$ a
  bounded open subset of $\mathbb{R}^{3}$, we assign the spectral
  projection $E_{\Delta}$ of the position operator for a particle in
  three dimensions.  Let $H(\mathbf{a})=0$ so that
  $U(t\mathbf{a})=e^{it0}=I$ for all $t\in \mathbb{R}$.  Since the
  energy in every state is zero, the energy condition is trivially
  satisfied.
  
  Note, however, that if the background spacetime is \emph{not}
  regarded as having a distinguished timelike direction, then this
  example violates the energy condition.  Indeed, the generator of an
  arbitrary timelike translation has the form
\begin{equation}
H(\mathbf{b}) \:=\: \mathbf{b}\cdot \mathbf{P}\:=
\:b_{0}0+b_{1}P_{1}+b_{2}P_{2}+b_{3}P_{3}\:=\:b_{1}P_{1}+b_{2}P_{2}+b_{3}P_{3},
\end{equation} where
$\mathbf{b}=(b_{0},b_{1},b_{2},b_{3})\in \mathbb{R}^{4}$ is a timelike
vector, and $P_{i}$ are the three orthogonal components of the total
momentum.  But since each $P_{i}$ has spectrum $\mathbb{R}$, the
spectrum of $H(\mathbf{b})$ is \emph{not} bounded from below when
$\mathbf{b}\neq \mathbf{a}$.  \hfill $\Box$
\end{example}

Malament's theorem does not require the full structure of Minkowski
spacetime (e.g., the Lorentz group).  Rather, it suffices to assume
that the affine space $M$ satisfies the following condition.
\begin{description}
\item[{\it No absolute velocity:}] Let $\mathbf{a}$ be a spacelike
  translation of $M$.  Then there is a pair $(\mathbf{b},\mathbf{c})$
  of timelike translations of $M$ such that
  $\mathbf{a}=\mathbf{b}-\mathbf{c}$.
\end{description}
Despite the fact that ``no absolute velocity'' is a feature of all
post-Galilean spacetimes, there are some who claim that the existence
of a (undetectable) preferred reference frame is perfectly consistent
with the empirical evidence on which relativistic theories are based
(cf.~Bell 1987, Chap.~9).  What is more, the existence of a preferred
frame is an absolutely essential feature of a number of ``realistic''
interpretations of quantum theory (cf.~Maudlin 1994, Chap.~7).  Thus,
this tacit assumption of Malament's theorem has the potential to be a
major source of contention for those wishing to maintain that there
can be a relativistic quantum mechanics of localizable particles.

There is a further worry about the generality of Malament's theorem:
It is not clear whether the result can be expected to hold for
arbitrary \emph{relativistic} spacetimes, or whether it is an artifact
of peculiar features of Minkowski spacetime (e.g., that space is
infinite).  To see this, suppose that $M$ is an arbitrary globally
hyperbolic manifold.  (That is, $M$ is a manifold that permits at
least one foliation $\cal{S}$ into spacelike hypersurfaces).  Although
$M$ will not typically have a translation group, we suppose that $M$
has a transitive Lie group $G$ of diffeomorphisms.  (Just as a
manifold is locally isomorphic to $\mathbb{R}^{n}$, a Lie group is
locally isomorphic to a group of translations.)  We require that $G$
has a representation $g\mapsto U(g)$ in the unitary operators on
$\hil{H}$; and, the translation covariance condition now says that
$E_{g(\Delta )}=U(g)E_{\Delta}U(g)^{*}$ for all $g\in G$.

The following example shows that Malament's theorem fails even for the
very simple case where $M$ is a two-dimensional cylinder.
\begin{example}  \label{cylinder} Let $M=\mathbb{R}\oplus S^{1}$, where $S^{1}$ is the
  one-dimensional unit circle, and let $G$ denote the Lie group of
  timelike translations and rotations of $M$.  It is not difficult to
  construct a unitary representation of $G$ that satisfies the energy
  bounded below condition.  (We can use the Hilbert space of
  square-integrable functions from $S^{1}$ into $\mathbb{C}$, and the
  procedure for constructing the unitary representation is directly
  analogous to the case of a single particle moving on a line.)  Fix a
  spacelike hypersurface $\Sigma$, and let $\mu$ denote the normalized
  rotation-invariant measure on $\Sigma$.  For each open subset
  $\Delta$ of $\Sigma$, let $E_{\Delta}=I$ if $\mu (\Delta )\geq 2/3$,
  and let $E_{\Delta}=0$ if $\mu (\Delta )<2/3$.  Then localizability
  holds, since for any pair $(\Delta ,\Delta ')$ of disjoint open
  subsets of $\Sigma$, either $\mu (\Delta )<2/3$ or $\mu (\Delta
  ')<2/3$.  \hfill $\Box$
\end{example}

Nonetheless, Examples~\ref{aristotle} and~\ref{cylinder} hardly serve
as physically interesting counterexamples to a strengthened version of
Malament's theorem.  In particular, in Example~\ref{aristotle} the
energy is identically zero, and therefore the probability for finding
the particle in a given region of space remains constant over time.
In Example~\ref{cylinder}, the particle is localized in every region
of space with volume greater than $2/3$, and the particle is never
localized in a region of space with volume less than $2/3$.  In the
following two sections, then, we will formulate explicit conditions to
rule out such pathologies, and we will use these conditions to derive
a strengthened version of Malament's theorem that applies to generic
spacetimes.

\section{Hegerfeldt's Theorem}
Hegerfeldt's (1998a, 1998b) recent results on localization apply to
arbitrary (globally hyperbolic) spacetimes, and they do not make us of
the ``no absolute velocity'' condition.  Thus, we will suppose
henceforth that $M$ is a globally hyperbolic spacetime, and we will
fix a foliation $\cal{S}$ of $M$, as well as a unique isomorphism
between any two hypersurfaces in this foliation.  If $\Sigma \in
\mathcal{S}$, we will write $\Sigma +t$ for the hypersurface that
results from ``moving $\Sigma$ forward in time by $t$ units''; and if
$\Delta$ is a subset of $\Sigma$, we will use $\Delta +t$ to denote
the corresponding subset of $\Sigma +t$.  We assume that there is a
representation $t\mapsto U_{t}$ of the time-translation group
$\mathbb{R}$ in the unitary operators on $\hil{H}$, and we will say
that the localization system $(\hil{H},\Delta \mapsto
E_{\Delta},t\mapsto U_{t})$ satisfies \emph{time-translation
  covariance} just in case $U_{t}E_{\Delta}U_{-t}=E_{\Delta +t}$ for
all $\Delta$ and all $t\in \mathbb{R}$.

Hegerfeldt's result is based on the following root lemma.
\begin{lemma}[Hegerfeldt] Suppose that $U_{t}=e^{itH}$, where $H$ is a
  self-adjoint operator with spectrum bounded from below.  Let $A$ be
  a positive operator (e.g., a projection operator).  Then for any
  state $\psi$, either
\[ \langle U_{t}\psi
,AU_{t}\psi \rangle \neq 0\,, \qquad \mbox{\textnormal{for almost all
    }}\; t\in \mathbb{R} ,\] or
\[ \hspace{-3em} \langle U_{t}\psi
,AU_{t}\psi \rangle =0\,, \qquad \mbox{\textnormal{for all }}\; t\in
\mathbb{R} \,.\] \label{gch}
\end{lemma}
Hegerfeldt claims that this lemma has the following consequence for
localization:
\begin{quote} If there exist particle states which are strictly
  localized in some finite region at $t=0$ and later move towards
  infinity, then finite propagation speed cannot hold for localization
  of particles. (Hegerfeldt 1998a, 243) \end{quote} Hegerfeldt's
argument for this conclusion is as follows: \begin{quote} Now, if the
  particle or system is strictly localized in $\Delta $ at $t=0$ it
  is, a fortiori, also strictly localized in any larger region $\Delta
  '$ containing $\Delta$.  If the boundaries of $\Delta '$ and
  $\Delta$ have a finite distance and \emph{if finite propagation
    speed holds} then the probability to find the system in $\Delta '$
  must also be $1$ for sufficiently small times, e.g.  $0\leq
  t<\epsilon$.  But then [Lemma~\ref{gch}], with $A\equiv I-E_{\Delta
    '}$, states that the system stays in $\Delta '$ for \emph{all}
  times.  Now, we can make $\Delta '$ smaller and let it approach
  $\Delta$.  Thus we conclude that if a particle or system is at time
  $t=0$ strictly localized in a region $\Delta$, then finite
  propagation speed implies that it stays in $\Delta$ for all times
  and therefore prohibits motion to infinity.  (Hegerfeldt 1998a,
  242--243; notation adapted, but italics in original)
\end{quote}  Let us attempt now to put this argument into a more precise form.

First, Hegerfeldt claims that the following is a consequence of
``finite propagation speed'': If $\Delta \subseteq \Delta '$, and if
the boundaries of $\Delta$ and $\Delta '$ have a finite distance, then
a state initially localized in $\Delta$ will continue to be localized
in $\Delta '$ for some finite amount of time.  We can capture this
precisely by means of the following condition.
\begin{description} 
\item[{\it No instantaneous wavepacket spreading (NIWS):}] If $\Delta
  \subseteq \Delta '$, and the boundaries of $\Delta$ and $\Delta '$
  have a finite distance, then there is an $\epsilon >0$ such that
  $E_{\Delta}\leq E_{\Delta '+t}$ whenever $0\leq t < \epsilon$.
\end{description} 
(Note that NIWS plus localizability entails strong causality.)  In the
argument, Hegerfeldt also assumes that if a particle is localized in
every one of a family of sets that ``approaches'' $\Delta$, then it is
localized in $\Delta$.  We can capture this assumption in the
following condition.
\begin{description}
\item[{\it Monotonicity:}] If $\{ \Delta _{n}:n \in \mathbb{N} \}$ is
  a downward nested family of subsets of $\Sigma$ such that $\bigcap
  _{n}\Delta _{n} =\Delta$, then $\bigwedge _{n}E_{\Delta
    _{n}}=E_{\Delta}$.  \end{description} Using this assumption,
Hegerfeldt argues that if NIWS holds, and if a particle is initially
localized in some finite region $\Delta$, then it will remain in
$\Delta$ for all subsequent times.  In other words, if $E_{\Delta}\psi
=\psi$, then $E_{\Delta}U_{t}\psi =U_{t}\psi$ for all $t\geq 0$.  We
can now translate this into the following rigorous no-go theorem.
\begin{thm*}[Hegerfeldt] Suppose that the localization system $(\hil{H},\Delta
  \mapsto E_{\Delta},t\mapsto U_{t})$ satisfies:
\begin{enumerate}
\item Monotonicity
\item Time-translation covariance
\item Energy bounded below
\item No instantaneous wavepacket spreading 
\end{enumerate}
Then $U_{t}E_{\Delta}U_{-t}=E_{\Delta}$ for all $\Delta \subset
\Sigma$ and all $t\in \mathbb{R}$. \label{reconstruct}
\end{thm*}
\noindent (For the proof of this theorem, see Appendix A.)  

Thus, conditions 1--4 can be satisfied only if the particle has
trivial dynamics.  If $M$ is an affine space, and if we add ``no
absolute velocity'' as a fifth condition in this theorem, then we get
the stronger conclusion that $E_{\Delta}=0$ for all bounded $\Delta$
(see Lemma~\ref{zowwy}, appendix).  Thus, there is an obvious
similarity between Hegerfeldt's and Malament's theorems.  However,
NIWS is a stronger causality assumption than microcausality.  In fact,
while NIWS plus localizability entails strong causality (and hence
microcausality), the following example shows that NIWS is not entailed
by the conjunction of strong causality, monotonicity, time-translation
covariance, and energy bounded below.
\begin{example} \label{descartes} Let $Q,P$ denote the standard
  position and momentum operators on $\hil{H}=L_{2}(\mathbb{R})$, and
  let $H=P^{2}/2m$ for some $m>0$.  Let $\Delta \mapsto
  E^{Q}_{\Delta}$ denote the spectral measure for $Q$.  Fix some
  bounded subset $\Delta _{0}$ of $\mathbb{R}$, and let
  $E_{\Delta}=E^{Q}_{\Delta}\otimes E^{Q}_{\Delta _{0}}$ (a projection
  operator on $\hil{H}\otimes \hil{H}$) for all Borel subsets $\Delta$
  of $\mathbb{R}$.  Thus, $\Delta \mapsto E_{\Delta}$ is a
  (non-normalized) projection-valued measure.  Let $U_{t}=I\otimes
  e^{itH}$, and let $E_{\Delta +t}=U_{t}E_{\Delta}U_{-t}$ for all
  $t\in \mathbb{R}$.  It is clear that monotonicity, time-translation
  covariance, and energy bounded below hold.  To see that strong
  causality holds, let $\Delta$ and $\Delta '$ be disjoint subsets of
  a single hyperplane $\Sigma$.  Then,
\begin{equation}
E_{\Delta}U_{t}E_{\Delta '}U_{-t}\:=\:E^{Q}_{\Delta}E^{Q}_{\Delta '}\otimes
  E^{Q}_{\Delta _{0}}E^{Q}_{\Delta _{0}+t}\:=\: 
0\otimes E^{Q}_{\Delta}E^{Q}_{\Delta _{0}+t}\:=\: 0 \,,\end{equation}
for all $t\in \mathbb{R}$.  On the other hand, 
  $U_{t}E_{\Delta}U_{-t}\neq E_{\Delta}$ for any nonempty $\Delta$ and
  for any $t\neq 0$.  
Thus, it follows from Hegerfeldt's theorem that NIWS fails.  \hfill $\Box$
\end{example}
Thus, we could not recapture the full strength of Malament's theorem
simply by adding ``no absolute velocity'' to the conditions of
Hegerfeldt's theorem.

\section{A Strengthened Hegerfeldt-Malament Theorem} \label{main}
Example~\ref{descartes} shows that Hegerfeldt's theorem fails if NIWS
is replaced by strong causality (or by microcausality).  On the other
hand, Example~\ref{descartes} is hardly a physically interesting
counterexample to a strengthened version of Hegerfeldt's theorem.  In
particular, if $\Sigma$ is a fixed spatial hypersurface, and if $\{
\Delta _{n}:n\in \mathbb{N} \}$ is a covering of $\Sigma$ by bounded
sets (i.e., $\bigcup _{n}\Delta _{n}=\Sigma$), then $\bigvee
_{n}E_{\Delta _{n}}=I\otimes E_{\Delta _{0}}\neq I\otimes I$.  Thus,
it is not certain that the particle will be detected \emph{somewhere
  or other} in space.  In fact, if $\{ \Delta _{n}:n\in \mathbb{N}\}$
is a covering of $\Sigma$ and $\{ \Pi _{n}:n\in \mathbb{N}\}$ is a
covering of $\Sigma +t$, then \begin{equation} \bigvee _{n\in
    \mathbb{N}}E_{\Delta_{n}}\:=\:I\otimes E_{\Delta _{0}}\:\neq \:
  I\otimes E_{\Delta _{0}+t}\:=\:\bigvee _{n\in \mathbb{N}}E_{\Pi
    _{n}} .\end{equation} Thus, the total probability for finding the
particle somewhere or other in space can change over time.

It would be completely reasonable to require that $\bigvee
_{n}E_{\Delta _{n}}=I$ whenever $\{ \Delta _{n}:n\in \mathbb{N}\}$ is
a covering of $\Sigma$.  This would be the case, for example, if the
mapping $\Delta \mapsto E_{\Delta}$ (restricted to subsets of
$\Sigma$) were the spectral measure of some position operator.
However, we propose that---at the very least---any physically
interesting model should satisfy the following weaker condition.
\begin{description}
\item[{\it Probability conservation:}] If $\{ \Delta _{n}:n\in
  \mathbb{N}\}$ is a covering of $\Sigma$, and $\{ \Pi _{n}:n\in
  \mathbb{N}\}$ is a covering of $\Sigma +t$, then $\bigvee
  _{n}E_{\Delta_{n}}=\bigvee _{n} E_{\Pi _{n}}$.  \end{description}
Probability conservation guarantees that there is a well-defined total
probability for finding the particle somewhere or other in space, and
this probability remains constant over time.  In particular, if both
$\{ \Delta _{n}:n\in \mathbb{N} \}$ and $\{ \Pi _{n}:n\in
\mathbb{N}\}$ consist of pairwise disjoint sets, then the
localizability condition entails that $\bigvee _{n}E_{\Delta
  _{n}}=\sum _{n}E_{\Delta _{n}}$ and $\bigvee _{n}E_{\Pi _{n}}=\sum
_{n}E_{\Pi _{n}}$.  In this case, probability conservation is
equivalent to
\begin{equation} \sum _{n\in \mathbb{N}}\mathrm{Prob} ^{\psi}(E_{\Delta
 _{n}})\:=\:\sum _{n\in \mathbb{N}}\mathrm{Prob}^{\psi}(E_{\Pi _{n}}) \,,\end{equation} 
for any state $\psi$.  Note, finally, that probability conservation is neutral with respect to 
relativistic and non-relativistic models.\footnote{Probability
  conservation would fail if a particle could escape to infinity in a 
finite amount of time (cf.~Earman
  1986, 33).  However, a particle can escape to infinity
only if there is an infinite potential well, and this would violate the energy 
condition.  Thus, given the energy condition, 
probability conservation should also hold for
non-relativistic particle theories.}   

\begin{thm}[Strengthened Hegerfeldt-Malament Theorem]  Suppose that
  the localization system $(\hil{H}, \Delta \mapsto
  E_{\Delta},t\mapsto U_{t})$ satisfies:
  \begin{enumerate}
  \item Localizability \label{ortho}
\item Probability conservation \label{conservation}
\item Time-translation covariance \label{covariance}
\item Energy bounded below \label{energy}
  \item Microcausality \label{micro}
\end{enumerate} Then $U_{t}E_{\Delta}U_{-t}=E_{\Delta}$ for all
$\Delta$ and all $t\in \mathbb{R}$.  \label{improve} \end{thm}
\noindent (For the proof of this theorem, see Appendix A.)  

If $M$ is an affine space, and if we add ``no absolute velocity'' as a
sixth condition in this theorem, then it follows that $E_{\Delta}=0$
for all $\Delta$ (see Lemma~\ref{zowwy}).  Thus, modulo the
probability conservation condition, Theorem~\ref{improve} recaptures
the full strength of Malament's theorem.  Moreover, we can now trace
the difficulties with localization to microcausality \emph{alone}:
there are localizable particles only if it is possible to have
act-outcome correlations at spacelike separation.

We now give examples to show that each condition in
Theorem~\ref{improve} is indispensable; that is, no four of the
conditions suffices to entail the conclusion.
(Example~\ref{aristotle} shows that conditions 1--5 can be
simultaneously satisfied.)  Suppose for simplicity that $M$ is
two-dimensional.  (All examples work in the four-dimensional case as
well.)  Let $Q,P$ be the standard position and momentum operators on
$L_{2}(\mathbb{R})$, and let $H=P^{2}/2m$.  Let $\Sigma$ be a spatial
hypersurface in $M$, and suppose that a coordinatization of $\Sigma$
has been fixed, so that there is a natural association between each
bounded open subset $\Delta$ of $\Sigma$ and a corresponding spectral
projection $E_{\Delta}$ of $Q$.
\begin{description}
\item[\textnormal{(1+2+3+4)}] (a) Consider the standard localization
  system for a single non-relativistic particle.  That is, let
  $\Sigma$ be a fixed spatial hyperplane, and let $\Delta \mapsto
  E_{\Delta}$ (with domain the Borel subsets of $\Sigma$) be the
  spectral measure for $Q$.  For $\Sigma +t$, set $E_{\Delta
    +t}=U_{t}E_{\Delta}U_{-t}$, where $U_{t}=e^{itH}$.  (b)~The
  Newton-Wigner approach to relativistic QM uses the standard
  localization system for a non-relativistic particle, only replacing
  the non-relativistic Hamiltonian $P^{2}/2m$ with the relativistic
  Hamiltonian $(P^{2}+m^{2}I)^{1/2}$, whose spectrum is also bounded
  from below.
\item[\textnormal{(1+2+3+5)}] (a) For a mathematically simple (but
  physically uninteresting) example, take the first example above and
  replace the Hamiltonian $P^{2}/2m$ with $P$.  In this case,
  microcausality trivially holds, since $U_{t}E_{\Delta}U_{-t}$ is
  just a shifted spectral projection of $Q$.  (b)~For a physically
  interesting example, consider the relativistic quantum theory of a
  single spin-$1/2$ electron (see Section~2).  Due to the negative
  energy solutions of the Dirac equation, the spectrum of the
  Hamiltonian is not bounded from below.
\item[\textnormal{(1+2+4+5)}] Consider the the standard localization
  system for a non-relativistic particle, but set $E_{\Delta
    +t}=E_{\Delta}$ for all $t\in \mathbb{R}$.  Thus, we escape the
  conclusion of trivial dynamics, but only by disconnecting the
  (nontrivial) unitary dynamics from the (trivial) association of
  projections with spatial regions.
\item[\textnormal{(1+3+4+5)}] (a) Let $\Delta _{0}$ be some bounded
  open subset of $\Sigma$, and let $E_{\Delta _{0}}$ be the
  corresponding spectral projection of $Q$.  When $\Delta \neq \Delta
  _{0}$, let $E_{\Delta}=0$.  Let $U_{t}=e^{itH}$, and let $E_{\Delta
    +t}=U_{t}E_{\Delta}U_{-t}$ for all $\Delta$.  This example is
  physically uninteresting, since the particle cannot be localized in
  any region besides $\Delta _{0}$, including proper supersets of
  $\Delta _{0}$.  (b)~See Example~\ref{descartes}.
\item[\textnormal{(2+3+4+5)}] Let $\Delta _{0}$ be some bounded open
  subset of $\Sigma$, and let $E_{\Delta _{0}}$ be the corresponding
  spectral projection of $Q$.  When $\Delta \neq \Delta _{0}$, let
  $E_{\Delta}=I$.  Let $U_{t}=e^{itH}$, and let $E_{\Delta
    +t}=U_{t}E_{\Delta}U_{-t}$ for all $\Delta$.  Thus, the particle
  is always localized in every region other than $\Delta _{0}$, and is
  sometimes localized in $\Delta _{0}$ as well.  \end{description}

\section{Are there Unsharply Localizable Particles?} \label{unsharp}
We have argued that attempts to undermine the four explicit
assumptions of Malament's theorem are unsuccessful.  We have also now
shown that the tacit assumption of ``no absolute velocity'' is not
necessary to derive Malament's conclusion.  And, yet, there is one
more loophole in the argument against a relativistic quantum mechanics
of localizable particles.  In particular, the basic assumption of a
family $\{ E_{\Delta} \}$ of localizing projections is unnecessary; it
is possible to have a quantum-mechanical particle theory in the
absence of localizing projections.  What is more, one might object to
the use of localizing projections on the grounds that they represent
an unphysical idealization---viz., that a ``particle'' can be
completely contained in a finite region of space with a sharp
boundary, when in fact it would require an infinite amount of energy
to prepare a particle in such a state.  Thus, there remains a
possibility that relativistic causality can be reconciled with
``unsharp'' localizability.

To see how we can define ``particle talk'' without having projection
operators, consider the relativistic theory of a single spin-$1/2$
electron (where we now restrict to the subspace
$\hil{H}_{\mathrm{pos}}$ of positive energy solutions of the Dirac
equation).  In order to treat the `$\mathib{x}$' of the Dirac
wavefunction as an observable, we need only to define a probability
amplitude and density for the particle to be found at $\mathib{x}$;
and these can be obtained from the Dirac wavefunction itself.  That
is, for a subset $\Delta$ of $\Sigma$, we set \begin{equation}
  \mathrm{Prob}^{\psi}(\mathib{x}\in \Delta)=\int _{\Delta}\abs{\psi
    (\mathib{x})}^{2}d\mathib{x} \,. \end{equation} Now let $\Delta
\mapsto E_{\Delta}$ be the spectral measure for the standard position
operator on the Hilbert space $\hil{H}$ (which includes both positive
and negative energy solutions).  That is, $E_{\Delta}$ multiplies a
wavefunction by the characteristic function of $\Delta$.  Let $F$
denote the orthogonal projection of $\hil{H}$ onto
$\hil{H}_{\mathrm{pos}}$.  Then,
\begin{equation} \int _{\Delta}\abs{\psi
    (\mathib{x})}^{2}d\mathib{x}=\langle \psi ,E_{\Delta}\psi
  \rangle =\langle \psi ,FE_{\Delta}\psi \rangle,\end{equation} 
for any $\psi \in \hil{H}_{\mathrm{pos}}$.  Thus, we can apply the standard recipe to
the operator $FE_{\Delta}$ (defined on $\hil{H}_{\mathrm{pos}}$)
to compute the probability that the particle will be found within
$\Delta$.  However, $FE_{\Delta}$ does \emph{not} define a projection operator
on $\hil{H}_{\mathrm{pos}}$.  (In fact, it can be shown that
$FE_{\Delta}$ does not have any eigenvectors with eigenvalue $1$.)
Thus, we do not need a family of \emph{projection} operators in order
to define probabilities for localization. 

Now, in general, to define the probability that a particle will be
found in $\Delta$, we need only assume that there is an operator
$A_{\Delta}$ such that $\langle \psi ,A_{\Delta}\psi \rangle \in
[0,1]$ for any unit vector $\psi$.  Such operators are called
\emph{effects}, and include the projection operators as a proper
subclass.  Thus, we say that the triple $(\hil{H},\Delta \mapsto
A_{\Delta},\mathbf{a}\mapsto U(\mathbf{a}))$ is an \emph{unsharp
  localization system} over $M$ just in case $\Delta \mapsto
A_{\Delta}$ is a mapping from subsets of hyperplanes in $M$ to effects
on $\hil{H}$, and $\mathbf{a}\mapsto U(\mathbf{a})$ is a continuous
representation of the translation group of $M$ in unitary operators on
$\hil{H}$.  (We assume for the present that $M$ is again an affine
space.)

Most of the conditions from the previous sections can be applied, with
minor changes, to unsharp localization systems.  In particular, since
the energy bounded below condition refers only to the unitary
representation, it can be carried over intact; and translation
covariance also generalizes straightforwardly.  However, we will need
to take more care with microcausality and with localizability.

If $E$ and $F$ are projection operators, $[E,F]=0$ just in case for
any state, the statistics of a measurement of $F$ are not affected by
a non-selective measurement of $E$ and vice versa (cf.~Malament 1996,
5).  This fact, along with the assumption that $E_{\Delta}$ is
measurable in $\Delta$, motivates the microcausality assumption.  For
the case of an association of arbitrary effects with spatial regions,
Busch (1999, Proposition 2) has shown that $[A_{\Delta},A_{\Delta
  '}]=0$ just in case for any state, the statistics for a measurement
of $A_{\Delta}$ are not affected by a non-selective measurement of
$A_{\Delta '}$ and vice versa.  Thus, we may carry over the
microcausality assumption intact, again seen as enforcing a
prohibition against act-outcome correlations at spacelike separation.

The localizability condition is motivated by the idea that a particle
cannot be simultaneously localized (with certainty) in two disjoint
regions of space.  In other words, if $\Delta$ and $\Delta '$ are
disjoint subsets of a single hyperplane, then $\langle \psi
,E_{\Delta}\psi \rangle =1$ entails that $\langle \psi ,E_{\Delta
  '}\psi \rangle =0$.  It is not difficult to see that this last
condition is equivalent to the assumption that $E_{\Delta}+E_{\Delta
  '}\leq I$.  That is, \begin{equation} \langle \psi
  ,(E_{\Delta}+E_{\Delta '})\psi \rangle \leq \langle \psi ,I\psi
  \rangle \, ,\end{equation} for any state $\psi$.  Now, it is an
accidental feature of projection operators (as opposed to arbitrary
effects) that $E_{\Delta}+E_{\Delta '}\leq I$ is equivalent to
$E_{\Delta}E_{\Delta '}=0$.  Thus, the apropriate generalization of
localizability to unsharp localization systems is the following
condition.
\begin{description}
\item[{\it Localizability:}] If $\Delta$ and $\Delta '$ are disjoint
  subsets of a single hyperplane, then \newline \mbox{$A_{\Delta}+A_{\Delta
      '}\leq I$}.
\end{description}
That is, the probability for finding the particle in $\Delta$, plus
the probability for finding the particle in some disjoint region
$\Delta '$, never totals more than $1$.  It would, in fact, be
reasonable to require a slightly stronger condition, viz., the
probability of finding a particle in $\Delta$ plus the probability of
finding a particle in $\Delta '$ equals the probability of finding a
particle in $\Delta \cup \Delta'$.  If this is true for all states
$\psi$, we have:
\begin{description}
\item[{\it Additivity:}] If $\Delta$ and $\Delta '$ are disjoint
  subsets of a single hyperplane, then \newline
  \mbox{$A_{\Delta}+A_{\Delta '}=A_{\Delta \cup \Delta '}$}.
\end{description}

With just these mild constraints, Busch (1999) was able to derive the
following no-go result.
\begin{thm*}[Busch] Suppose that the unsharp localization system 
  $(\hil{H},\Delta \mapsto A_{\Delta},\mathbf{a}\mapsto
  U(\mathbf{a}))$ satisfies localizability, translation covariance,
  energy bounded below, microcausality, and no absolute velocity.
  Then, for all $\Delta$, $A_{\Delta}$ has no eigenvector with
  eigenvalue $1$.
\end{thm*}
Thus, it is not possible for a particle to be localized with certainty
in any bounded region $\Delta$.  Busch's theorem, however, leaves it
open question whether there are (nontrivial) ``strongly unsharp''
localization systems that satisfy microcausality.  The following
result shows that there are not.

\begin{thm} Suppose that the unsharp localization system $(\hil{H},\Delta \mapsto
  A_{\Delta},\mathbf{a}\mapsto U(\mathbf{a}))$ satisfies:
  \begin{enumerate}
  \item Additivity 
  \item Translation covariance
  \item Energy bounded below
  \item Microcausality
  \item No absolute velocity
  \end{enumerate} Then $A_{\Delta}=0$ for all $\Delta$.  \label{pov}
\end{thm}
\noindent (For the proof of this theorem, see Appendix B.)  

Theorem~\ref{pov} shows that invoking the notion of unsharp
localization does nothing to resolve the tension between relativistic
causality and localizability.  For example, we can now show that the
(positive energy) Dirac theory---in which there are localizable
particles---violates relativistic causality.  Indeed, it is clear that
the conclusion of Theorem~\ref{pov} fails.\footnote{For any unit
  vector $\psi \in \hil{H}_{\mathrm{pos}}$, there is a bounded set
  $\Delta$ such that $\int _{\Delta}\abs{\psi}^{2}d\mathib{x}\neq 0$.
  Thus, $A_{\Delta}\neq 0$.}  On the other hand, additivity,
translation covariance, energy bounded below, and no absolute velocity
hold.  Thus, microcausality fails, and the (positive energy) Dirac
theory permits superluminal signalling.

Unfortunately, Theorem~\ref{pov} does not generalize to arbitrary
globally hyperbolic spacetimes, as the following example shows.
\begin{example} Let $M$ be the cylinder spacetime from
  Example~\ref{cylinder}.  Let $G$ denote the group of timelike
  translations and rotations of $M$, and let $g\mapsto U(g)$ be a
  positive energy representation of $G$ in the unitary operators on a
  Hilbert space $\hil{H}$.  For any $\Sigma \in \cal{S}$, let $\mu$
  denote the normalized rotation-invariant measure on $\Sigma$, and
  let $A_{\Delta}=\mu (\Delta )I$.  Then, conditions 1--5 of
  Theorem~\ref{pov} are satisfied, but the conclusion of the theorem
  is false.  \hfill $\Box$ \label{measure}
\end{example}

The previous counterexample can be excluded if we require there to be
a fixed positive constant $\delta$ such that, for each $\Delta$, there
is a state $\psi$ with $\langle \psi ,A_{\Delta}\psi \rangle \geq
\delta$.  In fact, with this condition added, Theorem~\ref{pov} holds
for any globally hyperbolic spacetime.  (The proof is an easy
modification of the proof we give in Appendix B.)  However, it is not
clear what physical motivation there could be for requiring this
further condition.  Note also that Example~\ref{measure} has trivial
dynamics; i.e., $U_{t}A_{\Delta}U_{-t}=A_{\Delta}$ for all $\Delta$.
We conjecture that every counterexample to a generalized version of
Theorem~\ref{pov} will have trivial dynamics.

Theorem~\ref{pov} strongly supports the conclusion that there is no
relativistic quantum mechanics of a single (localizable) particle; and
that the only consistent combination of special relativity and quantum
mechanics is in the context of quantum field theory.  However, neither
Theorem~\ref{improve} nor Theorem~\ref{pov} says anything about the
ontology of relativistic quantum field theory itself; they leave open
the possibility that relativistic quantum field theory might permit an
ontology of localizable particles.  To eliminate this latter
possibility, we will now proceed to present a more general result
which shows that there are no localizable particles in \emph{any}
relativistic quantum theory.

\section{Are there Localizable Particles in RQFT?} \label{numbers}
The localizability assumption is motivated by the idea that a
``particle'' cannot be detected in two disjoint spatial regions at
once.  However, in the case of a many-particle system, it is certainly
possible for there to be particles in disjoint spatial regions.  Thus,
the localizability condition does not apply to many-particle systems;
and Theorems~\ref{improve} and~\ref{pov} cannot be used to rule out a
relativistic quantum mechanics of $n>1$ localizable particles.

Still, one might argue that we could use $E_{\Delta}$ to represent the
proposition that a measurement is certain to find that \emph{all} $n$
particles lie within $\Delta$, in which case localizability should
hold.  Note, however, that when we alter the interpretation of the
localization operators $\{ E_{\Delta} \}$, we must alter our
interpretation of the conclusion.  In particular, the conclusion now
shows only that it is not possible for all $n$ particles to be
localized in a bounded region of space.  This leaves open the
possibility that there are localizable particles, but that they are
governed by some sort of ``exclusion principle'' that prohibits them
all from clustering in a bounded spacetime region.

Furthermore, Theorems~\ref{improve} and~\ref{pov} only show that it is
impossible to define \emph{position operators} that obey appropriate
relativistic constraints.  But it does not immediately follow from
this that we lack any notion of localization in relativistic quantum
theories.  Indeed,
\begin{quote} ...a position operator is
  inconsistent with relativity.  This compels us to find another way
  of modeling localization of events.  In field theory, we model
  localization by making the observables dependent on position in
  spacetime.  (Ticiatti 1999, 11) \end{quote} However, it is not a
peculiar feature of \emph{relativistic} quantum field theory that it
lacks a position operator: Any quantum field theory (either
relativistic or non-relativistic) will model localization by making
the observables dependent on position in spacetime.  Moreover, in the
case of non-relativistic QFT, these ``localized'' observables suffice
to provide us with a concept of localizable particles.  In particular,
for each spatial region $\Delta$, there is a ``number operator''
$N_{\Delta}$ whose eigenvalues give the number of particles within the
region $\Delta$.  Thus, we have no difficultly in talking about the
particle content in a given region of space despite the absence of any
position operator.

Abstractly, a number operator $N$ on $\hil{H}$ is any operator with
eigenvalues contained in $\{ 0,1,2,\dots \}$.  In order to describe
the number of particles locally, we require an association $\Delta
\mapsto N_{\Delta}$ of subsets of spatial hyperplanes in $M$ to number
operators on $\hil{H}$, where $N_{\Delta}$ represents the number of
particles in the spatial region $\Delta$.  If $\mathbf{a}\mapsto
U(\mathbf{a})$ is a unitary representation of the translation group,
we say that the triple $(\hil{H},\Delta \mapsto
N_{\Delta},\mathbf{a}\mapsto U(\mathbf{a}))$ is a \emph{system of
  local number operators} over $M$.  Note that a localization system
$\locsys$ is a special case of a system of local number operators
where the eigenvalues of each $N_{\Delta}$ are restricted to $\{ 0,1
\}$.  Furthermore, if we loosen our assumption that number operators
have a discrete spectrum, and instead require only that they have
spectrum contained in $[0,\infty )$, then we can also include unsharp
localization systems within the general category of systems of local
number operators.  Thus, a system of local number operators is the
\emph{minimal} requirement for a concept of localizable particles in
any quantum theory.

In addition to the natural analogues of the energy bounded below
condition, translation covariance, and microcausality, we will be
interested in the following two requirements on a system of local
number operators:\footnote{Due to the unboundedness of number
  operators, we would need to take some care in giving technically
  correct versions of the following conditions.  In particular, the
  additivity condition should technically include the clause that
  $N_{\Delta}$ and $N_{\Delta '}$ have a common dense domain, and the
  operator $N_{\Delta \cup \Delta '}$ should be thought of as the
  self-adjoint closure of $N_{\Delta}+N_{\Delta '}$.  In the number
  conservation condition, the sum $N=\sum _{n}N_{\Delta _{n}}$ can be
  made rigorous by exploiting the correspondence between self-adjoint
  operators and ``quadratic forms'' on $\hil{H}$.  In particular, we
  can think of $N$ as deriving from the upper bound of quadratic forms
  corresponding to finite sums.}
\begin{description}
\item[{\it Additivity:}] If $\Delta$ and $\Delta '$ are disjoint
  subsets of a single hyperplane, then \newline $N_{\Delta}+N_{\Delta
    '}=N_{\Delta \cup \Delta '}$.
\item[{\it Number conservation:}] If $\{ \Delta _{n}:n\in \mathbb{N}
  \}$ is a disjoint covering of $\Sigma$, then the sum $\sum
  _{n}N_{\Delta _{n}}$ converges to a densely defined, self-adjoint
  operator $N$ on $\hil{H}$ (independent of the chosen covering), and
  $U(\mathbf{a})NU(\mathbf{a})^{*}=N$ for any timelike translation
  $\mathbf{a}$ of $M$.
\end{description}
Additivity asserts that, when $\Delta$ and $\Delta '$ are disjoint,
the expectation value (in any state $\psi$) for the number of
particles in $\Delta \cup \Delta '$ is the sum of the expectations for
the number of particles in $\Delta$ and the number of particles in
$\Delta '$.  In the pure case, it asserts that the number of particles
in $\Delta \cup \Delta '$ is the sum of the number of particles in
$\Delta$ and the number of particles in $\Delta '$.  The ``number
conservation'' condition tells us that there is a well-defined total
number of particles (at a given time), and that the total number of
particles does not change over time.  This condition holds for any
non-interacting model of QFT.  

It is a well-known consequence of the Reeh-Schlieder theorem that
relativistic quantum field theories do not admit systems of local
number operators (cf.~Redhead 1995).  We will now derive the same
conclusion from strictly weaker assumptions.  In particular, we show
that microcausality is the \emph{only} specifically relativistic
assumption needed for this result.  The relativistic spectrum
condition---which requires that the spectrum of the four-momentum lie
in the forward light cone, and which is used in the proof of the
Reeh-Schlieder theorem---plays no role in our
proof.\footnote{Microcausality is not only sufficient, but also
  necessary for the proof that there are no local number operators.
  The Reeh-Schlieder theorem entails the cyclicity of the vacuum
  state.  But the cyclicity of the vacuum state alone does not entail
  that there are no local number operators; we must also assume
  microcausality (cf.~Halvorson 2001, Requardt 1982).}

\begin{thm} Suppose that the system $(\hil{H},\Delta \mapsto
  N_{\Delta},\mathbf{a}\mapsto U(\mathbf{a}))$ of local number
  operators satisfies:
 \begin{enumerate}
  \item Additivity
  \item Translation covariance \label{numbercovariance}
   \item Energy bounded below 
  \item Number conservation \label{numberconservation}
 \item Microcausality
 \item No absolute velocity
\end{enumerate}
Then $N_{\Delta}=0$ for all $\Delta$.  \label{rqft} \end{thm}
\noindent (For the proof of the theorem, see Appendix C.)  

Thus, in every state, there are no particles in any local region.
This serves as a \emph{reductio ad absurdum} for any notion of
localizable particles in a relativistic quantum theory.

Unfortunately, Theorem~\ref{rqft} is not the strongest result we could
hope for, since ``number conservation'' can only be expected to hold
in the (trivial) case of non-interacting fields.  However, we would
need a more general approach in order to deal with interacting
relativistic quantum fields, because (due to Haag's theorem;
cf.~Streater and Wightman, 2000, 163) their dynamics are not unitarily
implementable on a fixed Hilbert space.  On the other hand it would be
wrong to think of this as indicating a limitation on the generality of
our conclusion: Haag's theorem also entails that interacting models of
RQFT have no number operators---either global or local.\footnote{If a
  total number operator exists in a representation of the canonical
  commutation relations, then that representation is quasiequivalent
  to a free-field (Fock) representation (Chaiken 1968).  However,
  Haag's theorem entails that in relativistic theories,
  representations with nontrivial interactions are \emph{not}
  quasiequivalent to a free-field representation.}  Still, it would be
interesting to recover this conclusion (perhaps working in a more
general algebraic setting) without using the full strength of Haag's
assumptions.

\section{Particle Talk without Particle Ontology}
The results of the previous sections show that, insofar as we can
expect any relativistic quantum theory theory to satisfy a few basic
conditions, these theories do not admit (localizable) particles into
their ontology.  We also considered and rejected several arguments
which attempt to show that one (or more) of these conditions can be
jettisoned without doing violence to the theory of relativity or to
quantum mechanics.  Thus, we have yet to find a good reason to reject
one of the premises on which our argument against localizable
particles is based.  However, Segal (1964) and Barrett (2000) claim
that we have independent grounds for rejecting the conclusion; that
is, we have good reasons for believing that there \emph{are}
localizable particles.

The argument for localizable particles appears to be very simple: Our
experience shows us that objects (particles) occupy finite regions of
space.  But the reply to this argument is just as simple: These
experiences are illusory!  Although no object is strictly localized in
a bounded region of space, an object can be well-enough localized to
give the appearance to us (finite observers) that it is strictly
localized.  In fact, relativistic quantum field theory \emph{itself}
shows how the ``illusion'' of localizable particles can arise, and how
talk about localizable particles can be a useful fiction.

In order to assess the possibility of ``approximately localized''
objects in relativistic quantum field theory, we shall now pursue the
investigation in the framework of algebraic quantum field
theory.\footnote{For general information on algebraic quantum field
  theory, see (Haag 1992) and (Buchholz 2000).  For specific
  information on particle detectors and ``almost local'' observables,
  see Chapter 6 of (Haag 1992) and Section 4 of (Buchholz 2000).}
Here, one assumes that there is a correspondence $\hil{O} \mapsto
R(\hil{O})$ between bounded open subsets of $M$ and subalgebras of
observables on some Hilbert space $\hil{H}$.  Observables in
$R(\hil{O})$ are considered to be ``localized'' (i.e., measurable) in
$\cal{O}$.  Thus, if $\mathcal{O}$ and $\mathcal{O}'$ are spacelike
separated, we require that $[A,B]=0$ for any $A\in R(\mathcal{O})$ and
$B\in R(\mathcal{O}')$.  Furthermore, we assume that there is a
continuous representation $\mathbf{a}\mapsto U(\mathbf{a})$ of the
translation group of $M$ in unitary operators on $\hil{H}$, and that
there is a unique ``vacuum'' state $\Omega \in \hil{H}$ such that
$U(\mathbf{a})\Omega =\Omega$ for all $\mathbf{a}$.  This latter
condition entails that the vacuum appears the same to all observers,
and that it is the unique state of lowest energy.

In this context, a particle detector can be represented by an effect
$C$ such that $\langle \Omega ,C\Omega \rangle =0$.  That is, $C$
should register no particles in the vacuum state.  However, the
Reeh-Schlieder theorem entails that no positive local observable can
have zero expectation value in the vacuum state.  Thus, we again see
that (strictly speaking) it is impossible to detect particles by means
of local measurements; instead, we will have to think of particle
detections as ``approximately local'' measurements.

If we think of an observable as representing a measurement procedure
(or, more precisely, an equivalence class of measurement procedures),
then the norm distance $\norm{C-C'}$ between two observables gives a
quantitative measure of the physical similarity between the
corresponding procedures.  (In particular, if $\norm{C-C'}<\delta$,
then the expectation values of $C$ and $C'$ never differ by more than
$\delta$.)\footnote{Recall that $\norm{C-C'}$ is defined as the
  supremum of $\norm{(C-C')\psi }$ as $\psi$ runs through the unit
  vectors in $\hil{H}$.  It follows, then, from the Cauchy-Schwarz
  inequality that $\abs{\langle \psi ,(C-C')\psi \rangle }\leq
  \norm{C-C'}$ for any unit vector $\psi$.}  Moreover, in the case of
real-world measurements, the existence of measurement errors and
environmental noise make it impossible for us to determine precisely
which measurement procedure we have performed.  Thus, practically
speaking, we can at best determine a neighborhood of observables
corresponding to a concrete measurement procedure.

In the case of present interest, what we actually measure is always a
local observable---i.e., an element of $R(\mathcal{O})$, where
$\mathcal{O}$ is bounded.  However, given a fixed error bound
$\delta$, if an observable $C$ is within norm distance $\delta$ from
some local observable $C'\in R(\mathcal{O})$, then a measurement of
$C'$ will be practically indistinguishable from a measurement of $C$.
Thus, if we let \begin{equation} R_{\delta}(\mathcal{O})= \{ C:
  \exists C' \in R(\mathcal{O})\; \mbox{such that}\;
  \norm{C-C'}<\delta \} ,\end{equation} denote the family of
observables ``almost localized'' in $\mathcal{O}$, then `FAPP' (i.e.,
`for all practical purposes') we can locally measure any observable
from $R_{\delta}(\mathcal{O})$.  That is, measurement of an element
from $R_{\delta}(\mathcal{O})$ can be simulated to a high degree of
accuracy by local measurement of an element from $R(\mathcal{O})$.
However, for any local region $\cal{O}$, and for any $\delta >0$,
$R_{\delta}(\mathcal{O})$ \emph{does} contain (nontrivial) effects
that annihilate the vacuum.\footnote{Suppose that $A\in R(\hil{O})$,
  and let $A(\mathib{x})=U(\mathib{x})AU(\mathib{x})^{*}$.  If $f$ is
  a test function on $M$ whose Fourier transform is supported in the
  complement of the forward light cone, then $L=\int
  f(\mathib{x})A(\mathib{x})d\mathib{x}$ is almost localized in
  $\hil{O}$ and $\langle \Omega ,L\Omega \rangle=0$ (cf.~Buchholz
  2000, 7).}  Thus, particle detections can always be simulated by
purely local measurements; and the appearance of (fairly-well)
localized objects can be explained without the supposition that there
are localizable particles in the strict sense.

However, it may not be easy to pacify Segal and Barrett with a FAPP
solution to the problem of localization.  Both appear to think that
the absence of localizable particles (in the strict sense) is not
simply contrary to our manifest experience, but would undermine the
very possiblity of objective empirical science.  For example, Segal
claims that,
\begin{quote} ...it
  is an elementary fact, \emph{without which experimentation of the
    usual sort would not be possible}, that particles are indeed
  localized in space at a given time. (Segal 1965, 145; our italics)
\end{quote} Furthermore, ``particles would not be observable without their localization in
space at a particular time'' (1964, 139).  In other words,
experimentation involves observations of particles, and these
observations can occur only if particles are localized in space.
Unfortunately, Segal does not give any argument for these claims.  It
seems to us, however, that the moral we should draw from the no-go
theorems is that Segal's account of observation is false.  In
particular, it is not (strictly speaking) true that we observe
particles.  Rather, there are `observation events', and these
observation events are consistent (to a good degree of accuracy) with
the supposition that they are brought about by (localizable)
particles.

Like Segal, Barrett (2000) claims that we will have trouble explaining
how empirical science can work if there are no localizable particles.
In particular, Barrett claims that empirical science requires that we
be able to keep an account of our measurement results so that we can
compare these results with the predictions of our theories.
Furthermore, we identify measurement records by means of their
location in space.  Thus, if there were no localized objects, then
there would be no identifiable measurement records, and ``...it would
be difficult to account for the possibility of empirical science at
all'' (Barrett 2000, 3).

However, it's not clear what the difficulty here is supposed to be.
On the one hand, we have seen that relativistic quantum field theory
does predict that the appearances will be FAPP consistent with the
supposition that there are localized objects.  So, for example, we
could distinguish two record tokens at a given time if there were two
disjoint regions $\mathcal{O}$ and $\mathcal{O}'$ and particle
detector observables $C\in R_{\delta}(\mathcal{O})$ and $C'\in
R_{\delta}(\mathcal{O}')$ (approximated by observables \emph{strictly}
localized in $\mathcal{O}$ and $\mathcal{O}$ respectively) such that
$\langle \psi ,C\psi \rangle \approx 1$ and $\langle \psi ,C'\psi
\rangle \approx 1$.  Now, it may be that Barrett is also worried about
how, given a field ontology, we could assign any sort of
trans-temporal identity to our record tokens.  But this problem,
however important philosophically, is distinct from the problem of
localization.  Indeed, it also arises in the context of
non-relativistic quantum field theory, where there is \emph{no}
problem with describing localizable particles.  Finally, Barrett might
object that once we supply a quantum-theoretical model of a particle
detector itself, then the superposition principle will prevent the
field and detector from getting into a state where there is a fact of
the matter as to whether, ``a particle has been detected in the region
$\mathcal{O}$.''  But this is simply a restatement of the standard
quantum measurement problem that infects \emph{all} quantum
theories---and we have made no pretense of solving that here.

\section{Conclusion}
Malament claims that his theorem justifies the belief that,
\begin{quote}
  ...in the attempt to reconcile quantum mechanics with relativity
  theory...one is driven to a field theory; all talk about
  ``particles'' has to be understood, at least in principle, as talk
  about the properties of, and interactions among, quantized fields.
  (Malament 1996, 1) \end{quote} We have argued that the first claim
is correct---quantum mechanics and relativity can be reconciled only
in the context of quantum field theory.  In order, however, to close a
couple of loopholes in Malament's argument for this conclusion, we
provided two further results (Theorems~\ref{improve} and~\ref{pov})
which show that the conclusion continues to hold for generic
spacetimes, as well as for ``unsharp'' localization observables.  We
then went on to show that relativistic quantum field theory also does
not permit an ontology of localizable particles; and so, strictly
speaking, our talk about localizable particles is a fiction.
Nonetheless, relativistic quantum field theory does permit \emph{talk}
about particles---albeit, if we understand this talk as really being
about the properties of, and interactions among, quantized fields.
Indeed, modulo the standard quantum measurement problem, relativistic
quantum field theory has no trouble explaining the appearance of
macroscopically well-localized objects, and shows that our talk of
particles, though a \emph{fa{\c{c}}on de parler}, has a legitimate
role to play in empirically testing the theory.

\vspace{1em} {\it Acknowledgments:} We would like to thank Jeff
Barrett and David Malament for helpful correspondence.

\begin{appendix}
\section{Appendix}
\noindent {\bf Theorem (Hegerfeldt).}  {\it Suppose that the 
  localization system $(\hil{H},\Delta \mapsto E_{\Delta},t\mapsto
  U_{t})$ satisfies monotonicity, time-translation covariance, energy
  bounded below, and NIWS.  Then $U_{t}E_{\Delta}U_{-t}=E_{\Delta}$
  for all $\Delta \subset \Sigma$ and all $t\in \mathbb{R}$. }

\begin{proof} The formal proof corresponds directly to Hegerfeldt's
  informal proof.  Thus, let $\Delta$ be a subset of some spatial
  hypersurface $\Sigma$.  If $E_{\Delta}=0$ then obviously
  $U_{t}E_{\Delta}U_{-t}=E_{\Delta}$ for all $t\in \mathbb{R}$.  So,
  suppose that $E_{\Delta}\neq 0$, and let $\psi$ be a unit vector
  such that $E_{\Delta}\psi =\psi$.  Since $\Sigma$ is a manifold, and
  since $\Delta \neq \Sigma$, there is a family $\{ \Delta _{n}:n\in
  \mathbb{N} \}$ of subsets of $\Sigma$ such that, for each $n\in
  \mathbb{N}$, the distance between the boundaries of $\Delta _{n}$
  and $\Delta$ is nonzero, and such that $\bigcap _{n}\Delta
  _{n}=\Delta$.  Fix $n \in \mathbb{N}$.  By NIWS and time-translation
  covariance, there is an $\epsilon _{n}>0$ such that $E_{\Delta
    _{n}}U_{t}\psi = U_{t}\psi$ whenever $0\leq t< \epsilon _{n}$.
  That is, $\langle U_{t}\psi ,E_{\Delta _{n}}U_{t}\psi \rangle =1$
  whenever $0\leq t < \epsilon _{n}$.  Since energy is bounded from
  below, we may apply Lemma~\ref{gch} with $A=I-E_{\Delta _{n}}$ to
  conclude that $\langle U_{t}\psi ,E_{\Delta _{n}}U_{t}\psi \rangle
  =1$ for all $t\in \mathbb{R}$.  That is, $E_{\Delta _{n}}U_{t}\psi
  =U_{t}\psi$ for all $t\in \mathbb{R}$.  Since this holds for all
  $n\in \mathbb{N}$, and since (by monotonicity) $E_{\Delta}=\bigwedge
  _{n}E_{\Delta _{n}}$, it follows that $E_{\Delta}U_{t}\psi
  =U_{t}\psi$ for all $t\in \mathbb{R}$.  Thus, $U_{t}E_{\Delta
    }U_{-t}=E_{\Delta}$ for all $t\in \mathbb{R}$.
\end{proof}

\begin{lemma} Suppose that the localization system $(\hil{H},\Delta \mapsto
  E_{\Delta},\mathbf{a}\mapsto U(\mathbf{a}))$ satisfies
  localizability, time-translation covariance, and no absolute
  velocity.  Let $\Delta$ be a bounded spatial set.  If
  $U(\mathbf{a})E_{\Delta}U(\mathbf{a})^{*}=E_{\Delta}$ for all
  timelike translations $\mathbf{a}$ of $M$, then $E_{\Delta}=0$.
  \label{zowwy}
\end{lemma}

\begin{proof}  By no absolute velocity, there is a pair $(\mathbf{a},\mathbf{b})$
  of timelike translations such that $\Delta +(\mathbf{a}-\mathbf{b})$
  is in $\Sigma$ and is disjoint from $\Delta$.  By time-translation
  covariance, we have,
  \begin{equation} E_{\Delta
      +(\mathbf{a}-\mathbf{b})}\:=\:U(\mathbf{a})U(\mathbf{b})^{*}
    E_{\Delta}U(\mathbf{b})U(\mathbf{a})^{*}\:=\:E_{\Delta}
    .\end{equation} Thus, localizability entails that $E_{\Delta}$ is
  orthogonal to itself, and so $E_{\Delta}=0$.  \end{proof}

\begin{lemma} Let $\{ \Delta _{n}:n=0,1,2,\dots 
  \}$ be a covering of $\Sigma$, and let $E=\bigvee
  _{n=0}^{\infty}E_{\Delta _{n}}$.  If probability conservation and
  time-translation covariance hold, then $U_{t}EU_{-t}=E$ for all
  $t\in \mathbb{R}$.
  \label{commute} \end{lemma}

\begin{proof}  Since $\{ \Delta _{n}+t:n\in \mathbb{N} \}$ is a covering of $\Sigma +{t}$,
  probability conservation entails that $\bigvee _{n}E_{\Delta
    _{n}+t}=E$.  Thus,
\begin{eqnarray}
U_{t}EU_{-t} &=& U_{t}\biggl[ \,\bigvee _{n=0}^{\infty} 
E_{\Delta _{n}} \,\biggr] U_{-t} \:=\: 
\bigvee _{n=0}^{\infty}\,\biggl[ U_{t}E_{\Delta
    _{n}}U_{-t} \biggr] \\
&=& \bigvee _{n=0}^{\infty}E_{\Delta _{n}+t} \: =\:
    E, \end{eqnarray} where the third equality follows from
    time-translation covariance.  \end{proof}

In order to prove the next result, we will need to invoke the
following lemma from Borchers (1967).
\begin{lemma*}[Borchers] Let $U_{t}=e^{itH}$, where $H$ is a self-adjoint
  operator with spectrum bounded from below.  Let $E$ and $F$ be
  projection operators such that $EF=0$.  If there is an $\epsilon >0$
  such that \[ [E,U_{t}FU_{-t}]=0 \,, \qquad 0\leq t<\epsilon \,,\]
  then $EU_{t}FU_{-t}=0$ for all $t\in \mathbb{R}$.
\end{lemma*}

\begin{lemma} Let $U_{t}=e^{itH}$,
  where $H$ is a self-adjoint operator with spectrum bounded from
  below.  Let $\{ E_{n}:n=0,1,2,\dots \}$ be a family of projection
  operators such that $E_{0}E_{n}=0$ for all $n\geq 1$, and let
  $E=\bigvee _{n=0}^{\infty}E_{n}$.  If $U_{t}EU_{-t}=E$ for all $t\in
  \mathbb{R}$, and if for each $n\geq 1$ there is an $\epsilon _{n}>0$
  such that
\begin{equation} [E_{0},U_{t}E_{n}U_{-t}]=0 , \qquad 
0\leq t<\epsilon _{n} , \label{wiggle} \end{equation} then $U_{t}E_{0}U_{-t}=E_{0}$ for all
$t\in \mathbb{R}$.
\label{root} \end{lemma}

\begin{proof} If $E_{0}=0$ then the conclusion obviously
  holds.  Suppose then that $E_{0}\neq 0$, and let $\psi$ be a unit
  vector in the range of $E_{0}$.  Fix $n\geq 1$.
  Using~(\ref{wiggle}) and Borchers' lemma, it follows that
  $E_{0}U_{t}E_{n}U_{-t}=0$ for all $t\in \mathbb{R}$.  Then,
\begin{eqnarray}
\| E_{n}U_{-t}\psi \| ^{2} &=&\langle U_{-t}\psi ,E_{n}U_{-t}\psi \rangle \:=\: \langle \psi
,U_{t}E_{n}U_{-t}\psi \rangle \\
&=& \langle E_{0}\psi
,U_{t}E_{n}U_{-t}\psi \rangle \:=\: \langle \psi
,E_{0}U_{t}E_{n}U_{-t}\psi \rangle 
\:=\: 0\, ,\end{eqnarray}
for all $t\in \mathbb{R}$.  Thus, $E_{n}U_{-t}\psi =0$ for all $n\geq 1$, and consequently, 
$[\,\bigvee _{n\geq 1}E_{n}]U_{-t}\psi =0$.  Since $E_{0}=E-[\, \bigvee
_{n\geq 1}E_{n}]$, and since (by assumption) $EU_{-t}=U_{-t}E$, it follows that \begin{equation}
E_{0}U_{-t}\psi =EU_{-t}\psi =U_{-t}E\psi =U_{-t}\psi \,, \end{equation}
for all $t\in \mathbb{R}$.  \end{proof}

\vspace{0.5em} \noindent {\bf Theorem~\ref{improve}.}  {\it Suppose
  that the localization system $(\hil{H}, \Delta \mapsto
  E_{\Delta},t\mapsto U_{t})$ satisfies localizability, probability
  conservation, time-translation covariance, energy bounded below, and
  microcausality.  Then $U_{t}E_{\Delta}U_{-t}=E_{\Delta}$ for all
  $\Delta$ and all $t\in \mathbb{R}$.}

\begin{proof} Let $\Delta$ be an open subset of $\Sigma$.  
  If $\Delta =\Sigma$ then probability conservation and
  time-translation covariance entail that $E_{\Delta}=E_{\Delta
    +t}=U_{t}E_{\Delta}U_{-t}$ for all $t\in \mathbb{R}$.  If $\Delta
  \neq \Sigma$ then, since $\Sigma$ is a manifold, there is a covering
  $\{ \Delta _{n}:n\in \mathbb{N} \}$ of $\Sigma \backslash \Delta$
  such that the distance between $\Delta _{n}$ and $\Delta $ is
  nonzero for all $n$.  Let $E_{0}=E_{\Delta}$, and let
  $E_{n}=E_{\Delta _{n}}$ for $n\geq 1$.  Then~\ref{ortho} entails
  that $E_{0}E_{n}=0$ when $n\geq 1$.  If we let $E=\bigvee
  _{n=0}^{\infty}E_{n}$ then probability conservation entails that
  $U_{t}EU_{-t}=E$ for all $t\in \mathbb{R}$ (see
  Lemma~\ref{commute}).  By time-translation covariance and
  microcausality, for each $n\geq 1$ there is an $\epsilon _{n}>0$
  such that
\begin{equation} [E_{0},U_{t}E_{n}U_{-t}]=0 , \qquad 0\leq t< \,\epsilon
  _{n}.
\end{equation}  Since the energy is bounded from below, Lemma~\ref{root} entails that 
$U_{t}E_{0}U_{-t}=E_{0}$ for all $t\in \mathbb{R}$.  That is,
$U_{t}E_{\Delta}U_{-t}=E_{\Delta}$ for all $t\in \mathbb{R}$.
\end{proof}

\section{Appendix}
\noindent {\bf Theorem~\ref{pov}.} {\it Suppose that the unsharp
  localization system $(\hil{H},\Delta \mapsto
  A_{\Delta},\mathbf{a}\mapsto U(\mathbf{a}))$ satisfies additivity,
  translation covariance, energy bounded below, microcausality, and no
  absolute velocity.  Then $A_{\Delta}=0$ for all $\Delta$.  }

\begin{proof} We prove by induction that $\norm{A_{\Delta}}\leq
  (2/3)^{m}$, for each $m\in \mathbb{N}$, and for each bounded
  $\Delta$.  For this, let $F_{\Delta}$ denote the spectral measure
  for $A_{\Delta}$.
  
  (Base case: $m=1$) Let $E_{\Delta}=F_{\Delta }(2/3,1)$.  We verify
  that $(\hil{H},\Delta \mapsto E_{\Delta},\mathbf{a}\mapsto
  U(\mathbf{a}))$ satisfies the conditions of Malament's theorem.
  Clearly, no absolute velocity and energy bounded below hold.
  Moreover, since unitary transformations preserve spectral
  decompositions, translation covariance holds; and since spectral
  projections of compatible operators are also compatible,
  microcausality holds.  To see that localizability holds, let
  $\Delta$ and $\Delta '$ be disjoint bounded subsets of a single
  hyperplane.  Then microcausality entails that $[A_{\Delta},A_{\Delta
    '}]=0$, and therefore $E_{\Delta}E_{\Delta '}$ is a projection
  operator.  Suppose for reductio ad absurdum that $\psi$ is a unit
  vector in the range of $E_{\Delta}E_{\Delta '}$.  By additivity,
  $A_{\Delta \cup \Delta '}=A_{\Delta}+A_{\Delta '}$, and we therefore
  obtain the contradiction:
\begin{equation}
1 \:\geq \:\langle \psi,A_{\Delta \cup \Delta '}\psi \rangle
\: = \: \langle \psi ,A_{\Delta}\psi \rangle +\langle \psi ,A_{\Delta
  '}\psi \rangle \:\geq \: 2/3+2/3 \,.\end{equation}
Thus, $E_{\Delta}E_{\Delta '}=0$, and Malament's theorem entails
that $E_{\Delta }=0$ for all $\Delta$.  Therefore,
  $A_{\Delta}=A_{\Delta}F_{\Delta }(0,2/3)$ has spectrum lying in
  $[0,2/3]$, and $\norm{A_{\Delta}}\leq 2/3$ for all bounded $\Delta$.
  
  (Inductive step) Suppose that $\norm{A_{\Delta}}\leq (2/3)^{m-1}$
  for all bounded $\Delta$.  Let $E_{\Delta}=F_{\Delta} (
  (2/3)^{m},(2/3)^{m-1})$.  In order to see that Malament's theorem
  applies to $(\hil{H},\Delta \mapsto E_{\Delta},\mathbf{a}\mapsto
  U(\mathbf{a}))$, we need only check that localizability holds.  For
  this, suppose that $\Delta$ and $\Delta '$ are disjoint subsets of a
  single hyperplane.  By microcausality, $[A_{\Delta},A_{\Delta
    '}]=0$, and therefore $E_{\Delta}E_{\Delta '}$ is a projection
  operator.  Suppose for reductio ad absurdum that $\psi$ is a unit
  vector in the range of $E_{\Delta}E_{\Delta '}$.  Since $\Delta \cup
  \Delta '$ is bounded, the induction hypothesis entails that
  $\norm{A_{\Delta \cup \Delta '}}\leq (2/3)^{m-1}$.  By additivity,
  $A_{\Delta \cup \Delta '}=A_{\Delta}+A_{\Delta '}$, and therefore we
  obtain the contradiction:
\begin{equation}
(2/3)^{m-1} \:\geq \:\langle \psi,A_{\Delta \cup \Delta '}\psi \rangle
\: = \: \langle \psi ,A_{\Delta}\psi \rangle +\langle \psi ,A_{\Delta '}\psi \rangle 
\:\geq \: (2/3)^{m}+(2/3)^{m} \,.\end{equation}
Thus, $E_{\Delta}E_{\Delta '}=0$, and Malament's
theorem entails that $E_{\Delta }=0$ for all $\Delta$.  
Therefore, $\norm{A_{\Delta}}\leq (2/3)^{m}$ for all bounded $\Delta$.  \end{proof}

\section{Appendix}
\vspace{1em} \noindent {\bf Theorem~\ref{rqft}.}  {\it Suppose that
  the system $(\hil{H},\Delta \mapsto N_{\Delta},\mathbf{a}\mapsto
  U(\mathbf{a}))$ of local number operators satisfies additivity,
  translation covariance, energy bounded below, number conservation,
  microcausality, and no absolute velocity.  Then, $N_{\Delta }=0$ for
  all bounded $\Delta$.}

\begin{proof} Let $N$ be the unique total number operator obtained
  from taking the sum $\sum _{n}N_{\Delta _{n}}$ where $\{ \Delta
  _{n}:n\in \mathbb{N}\}$ is a disjoint covering of $\Sigma$.  Note
  that for any $\Delta \subseteq \Sigma$, we can choose a covering
  containing $\Delta$, and hence, $N=N_{\Delta}+A$, where $A$ is a
  positive operator.  By microcausality, $[N_{\Delta},A]=0$, and
  therefore $[N_{\Delta},N]=[N_{\Delta},N_{\Delta}+A]=0$.
  Furthermore, for any vector $\psi$ in the domain of $N$, $\langle
  \psi ,N_{\Delta}\psi \rangle \leq \langle \psi ,N\psi \rangle$.
  
  Let $E$ be the spectral measure for $N$, and let $E_{n}=E(0,n)$.
  Then, $NE_{n}$ is a bounded operator with norm at most $n$.  Since
  $[E_{n},N_{\Delta}]=0$, it follows that \begin{equation} \langle
    \psi ,N_{\Delta}E_{n}\psi \rangle = \langle E_{n}\psi
    ,N_{\Delta}E_{n}\psi \rangle \leq \langle E_{n}\psi ,NE_{n}\psi
    \rangle \leq n \, ,\end{equation} for any unit vector $\psi$.
  Thus, $\norm{N_{\Delta}E_{n}}\leq n$.  Since $\bigcup
  _{n=1}^{\infty}E_{n}(\hil{H})$ is dense in $\hil{H}$, and since
  $E_{n}(\hil{H})$ is in the domain of $N_{\Delta}$ (for all $n$), it
  follows that if $N_{\Delta}E_{n}=0$, for all $n$, then
  $N_{\Delta}=0$.  We now concentrate on proving the antecedent.
  
  For each $\Delta$, let $A_{\Delta}=(1/n)N_{\Delta}E_{n}$.  We show
  that the structure $(\hil{H},\Delta \mapsto
  A_{\Delta},\mathbf{a}\mapsto U(\mathbf{a}))$ satisfies the
  conditions of Theorem~\ref{pov}.  Clearly, energy bounded below and
  no absolute velocity hold.  It is also straightforward to verify
  that additivity and microcausality hold.  To check translation
  covariance, we compute:
\begin{eqnarray}
\hspace{-2em} U(\mathbf{a})A_{\Delta}U(\mathbf{a})^{*} &=&
U(\mathbf{a})N_{\Delta}E_{n}U(\mathbf{a})^{*} \:=\: 
U(\mathbf{a})N_{\Delta}U(\mathbf{a})^{*}U(\mathbf{a})E_{n}U(\mathbf{a})^{*}
\\
&=& U(\mathbf{a})N_{\Delta}U(\mathbf{a})^{*}E_{n} \:=\: N_{\Delta
  +\mathbf{a}}E_{n} 
\:=\: A_{\Delta +\mathbf{a}}. \end{eqnarray}  
The third equality follows from number conservation, and the
fourth equality follows from translation covariance.
Thus, $N_{\Delta}E_{n}=A_{\Delta}=0$ for all $\Delta$.  Since this
holds for all $n\in \mathbb{N}$, $N_{\Delta}=0$ for
all $\Delta$.  \end{proof}

\end{appendix}

\newpage
\begin{center} {\sc references } \end{center}

Barrett, Jeffrey A. (2000), ``On the nature of
measurement records in relativistic quantum field theory'',
manuscript.

Bell, John S. (1987), {\it Speakable and Unspeakable in Quantum
  Mechanics}.  New York: Cambridge University Press.

Borchers, H.-J. (1967), ``A remark on a theorem of B.  Misra'', {\it
  Communications in Mathematical Physics} 4: 315--323.

Buchholz, Detlev (2000), ``Algebraic quantum field theory: A status
report'', math-ph/0011044.
  
Busch, Paul (1999), ``Unsharp localization and causality in
relativistic quantum theory'', {\it Journal of Physics A} 32:
6535--6546.

Chaiken, Jan M. (1968), ``Number operators for representations of the
canonical commutation relations'', {\it Communications in Mathematical
  Physics} 8: 164--184.

Dickson, W. Michael (1998), {\it Quantum Chance and Nonlocality.} New
York: Cambridge University Press.

Earman, John (1986), {\it A Primer on Determinism}. Boston: D.
Reidel.

Fleming, Gordon, and Jeremy Butterfield (1999), ``Strange positions'',
in J. Butterfield and C. Pagonis (eds.), {\it From Physics to
  Philosophy}. NY: Cambridge University Press, 108--165.

Haag, Rudolf (1992), {\it Local Quantum Physics}. New York: Springer.

Halvorson, Hans (2001), ``Reeh-Schlieder defeats Newton-Wigner: On
alternative localization schemes in relativistic quantum field
theory'', {\it Philosophy of Science}, forthcoming.

Hegerfeldt, Gerhard C. (1998a), ``Causality, particle localization and
positivity of the energy'', in A.  B{\"o}hm, et al. (eds.), {\it
  Irreversibility and Causality}. New York: Springer, 238--245.

Hegerfeldt, Gerhard C. (1998b), ``Instantaneous spreading and Einstein
causality in quantum theory'', {\it Annalen der Physik} 7: 716--725.

Holland, Peter R. (1993), {\it The Quantum Theory of Motion}. New
York: Cambridge University Press.

Malament, David (1996), ``In defense of dogma: Why there cannot be a
relativistic quantum mechanics of (localizable) particles'', in Rob
Clifton (ed.), {\it Perspectives on Quantum Reality}. Dordrecht:
Kluwer, 1--10.

Maudlin, Tim (1994), {\it Quantum Non-Locality and Relativity}.
Cambridge: Blackwell.

Redhead, Michael (1995), ``The vacuum in relativistic quantum field
theory'', in David Hull, Micky Forbes, and Richard M. Burian (eds.),
{\it PSA 1994}, v. 2. East Lansing, MI: Philosophy of Science
Association, 77--87.

Requardt, Manfred (1982), ``Spectrum condition, analyticity,
Reeh-Schlieder and cluster properties in non-relativistic
Galilei-invariant quantum theory'', {\it Journal of Physics A} 15:
3715--3723.

Schlieder, S. (1971), ``Zum kausalen Verhalten eines relativistischen
quantenmechanischen System'', in S.P. D{\"u}rr (ed.), {\it Quanten und
  Felder}.  Braunschweig: Vieweg, 145--160.

Segal, Irving E. (1964), ``Quantum fields and analysis in the solution
manifolds of differential equations'', in William T. Martin and Irving
E. Segal, (eds.), {\it Proceedings of a Conference on the Theory and
  Applications of Analysis in Function Space}.  Cambridge: MIT Press,
129--153.

Streater, Raymond F. and Arthur S. Wightman (2000), {\it PCT, Spin and
  Statistics, and All That}. Princeton: Princeton University Press.

Thaller, Bernd (1992), {\it The Dirac Equation}.  New York: Springer.

Ticiatti, Robin (1999), {\it Quantum Field Theory for Mathematicians}.
New York: Cambridge University Press.

\end{document}